\documentclass[hyper]{JHEP} 

\usepackage{epsfig}

\newcommand\fverb{\setbox\pippobox=\hbox\bgroup\verb}
\newcommand\fverbdo{\egroup\medskip\noindent%
			\fbox{\unhbox\pippobox}\ }
\newcommand\fverbit{\egroup\item[\fbox{\unhbox\pippobox}]}
\newbox\pippobox
\title{D-Brane Effective Actions and Particle Production
near the Beginning of  Tachyon Condensation}
\author{by J. Kluso\v{n}\\
	 Department of Theoretical Physics and Astrophysics\\
                   Faculty of Science, Masaryk University\\
Kotl\'{a}\v{r}sk\'{a} 2, 611 37, Brno\\
Czech Republic\\
	E-mail: \email{klu@physics.muni.cz}}

\preprint{\hepth{0312086}}

\abstract{In this paper we will study the
quantum field theory of fluctuation modes around
the classical solution that describes 
 tachyon condensation on unstable D-brane. 
We will calculate the number of
particle produced at times near 
the beginning of the rolling tachyon process.
  We will perform this calculation
for different tachyon effective actions and we will
find that the rate of particle production strongly 
depends  on the 
form of the effective action used for the 
description of the early
stage of the tachyon condensation.}

\def\ket #1{\left|#1\right>}

\def\bk{\mathbf{k}}
\def\bx{\mathbf{x}}
\def\by{\mathbf{y}}
\def\bz{\mathbf{z}}
\def\bb{\mathbf{B}}

\def\ss{\sin \frac{\tau}{\sqrt{2}}}
\def\st{\sinh \frac{\tau}{\sqrt{2}}}
\def\st2{\sinh^2 \frac{\tau}{\sqrt{2}}}
\def\ss2{\sin^2 \frac{\tau}{\sqrt{2}}}

\def\ik{\frac{d^p\bk}{(2\pi)^p}}
\def\tg{\tilde{G}}
\def\ma{\mathcal{A}}

\def\hphi{\hat{\phi}}
\def\hpi{\hat{\Pi}}
\def\omk{\omega_{0 \bk}}

\begin{document}
\section{Introduction}\label{first}
One of the most intriguing problems in string theory
is  understanding tachyon condensation and its
description in terms of tachyon effective theory. 
Generally, it seems to be difficult task to find
effective action for the tachyon field only since
the scale of masses of an infinite set of massive
string modes is the same as that of the tachyon mass.
Consequently keeping the tachyon and integrating
out all other string modes seems to be not
well justified. On the other hand we can still
believe that in certain situations (like non-BPS
D-brane decay, forming of the tachyon kinks and
vertex) some aspects of string dynamics can be described
by an effective field theory action involving 
only tachyon field and massless modes where all
other massive modes effectively decouples at the
vicinity of certain conformal points.

In recent years  there were many papers that discussed 
various forms  of  tachyon effective actions
\cite{Sen:2003tm,Sen:2002qa,Kutasov:2003er,
Sen:2003bc,Lambert:2002hk,
Sen:1999md,Garousi:2000tr,Bergshoeff:2000dq,
Kluson:2000iy,Minahan:2000ff,Minahan:2000tf,
Lambert:2003zr,Kim:2003in,
Garousi:2003ai,Kluson:2003sr,
Smedback:2003ur}. Explanation, why there are so many
proposals for tachyon effective actions 
was given recently in very interesting paper
\cite{Fotopoulos:2003yt}. As was stressed there
the form of the tachyon effective action may depend
on a choice of a region in field space where
it should be valid. One such a region corresponds to
the neighbourhood of the perturbative string vacuum
$T=0$ where we can reconstruct tachyon effective
action from string $S$-matrix.  Then we can expect
the effective action  
\begin{equation}\label{Tperturb}
S=-\int dt d^p\bx\left(\frac{1}{2}\eta^{\mu\nu}
\partial_{\mu}T\partial_{\nu}T-\frac{\mu^2}{2}T^2+
g_1T^4\dots\right) \ .
\end{equation}
Another possibility how to get tachyon effective
action is to  study tachyon dynamics near  end point
of tachyon evolution $T\gg 1$ where tachyon gets 
frozen. Then we can find from the derivative
expansion of the superstring partition function  on
the disc \cite{Tseytlin:2000mt}
\begin{equation}\label{Tlarge}
S=-\int dt d^p\bx e^{-\frac{1}{4}T^2}
\left[1+\frac{1}{2}(1+b_1T^2)\eta^{\mu\nu}
\partial_{\mu}T\partial_{\nu}T+\dots\right] \ , \
b_1=\ln 2-\frac{1}{2} \ .
\end{equation}
One may hope that this action may be used to
interpolate between  the regions of small $T$ 
(vicinity of the tachyon vacuum) and large $T$ where
we can expect that tachyon and other open string
modes are frozen. 
Another example of the tachyon effective action
is ``tachyon DBI'' Lagrangian
\begin{equation}\label{tachDBI}
S=-\int dt d^p\bx V(T)\sqrt{
-\det(\eta_{\mu\nu}+\partial_{\mu}T
\partial_{\nu}T)} \ .
\end{equation}
As was argued recently in 
\cite{Fotopoulos:2003yt} the meaning of
(\ref{tachDBI}) is not completely clear since
there is no reason to expect that higher-dimensional
derivative terms omitted in (\ref{tachDBI}) 
should be small on solutions of resulting 
equations.

We can also try to reconstruct the tachyon
effective action at a vicinity of 
other exact conformal points. One such 
a conformal point is 
time-dependent
background that represents exact boundary 
conformal field theory 
\cite{Sen:2002an,Sen:2002in,Sen:2002nu,Gutperle:2003xf}
\begin{equation}
T=f_0 e^{\mu x^0}+\overline{f}_0
e^{-\mu x^0} \ , \ 
\mu^2_{bose}=1 \ , \  \mu^2_{super}=\frac{1}{2} \ 
\end{equation}
or its special case ``rolling tachyon'' background
\begin{equation}\label{rol}
T=f_0e^{\mu x^0} \ .
\end{equation}
By demanding that generic first-derivative
Lagrangian
\footnote{We are using the convention
where $\eta_{\mu\nu}
=\mathrm{diag}(-1,1,\dots,1) \ , \mu, \nu=0,\dots, p \ ,
a,b=1,\dots,p \ , x^0=t \ ,
 \bx=(x^1,\dots,x^p) \ $
 and the fundamental
string tension has been set equal to
 $(2\pi)^{-1} \ , \alpha'=1$.}
\begin{equation}\label{lagK}
L=-V(T)K(\partial T) 
\end{equation}
should have (\ref{rol})(with
$\mu^2=\frac{1}{2}$ in the superstring 
case) as its exact solution fixes its
time-dependent part to be
\cite{Kutasov:2003er}
\footnote{For similar approach to this
problem in the bosonic case, see recent
paper \cite{Smedback:2003ur}.}
\begin{equation}\label{kutt}
L=-\frac{1}{1+\frac{T^2}{2}}\sqrt{1+
\frac{T^2}{2}-(\partial_0 T)^2} \ .
\end{equation}
If we now assume that (\ref{kutt}) has
direct Lorenz-covariant generalisation
we obtain the tachyon effective action
\begin{equation}\label{Kutact}
S=-\int dt d^p\bx\frac{1}{1+\frac{T^2}{2}}
\sqrt{1+\frac{T^2}{2}+\eta^{\mu\nu}
\partial_{\mu}T\partial_{\nu}T} \ .
\end{equation}
The action (\ref{Kutact}) after field 
redefinition $\frac{T}{\sqrt{2}}=
\sinh \frac{\tilde{T}}{\sqrt{2}}$ becomes
\begin{equation}\label{tachDBIx}
S=-\int dt d^p\bx V(\tilde{T})
\sqrt{1+\eta^{\mu\nu}
\partial_{\mu}\tilde{T}
\partial_{\nu}\tilde{T}} \   ,
V(\tilde{T})=\cosh\frac{\tilde{T}}{\sqrt{2}}
 \ . 
\end{equation}
which has the form of ``DBI'' 
tachyon action  (\ref{tachDBI}).

As  it is clear from previous
remarks 
it is very important to know where
 in the field theory space 
different tachyon effective actions 
 correctly describe the tachyon
dynamics. 
For example,  even if the rolling tachyon solution
$T=e^{\frac{x_0}{\sqrt{2}}}$ is an
 exact solution of the equation
of motion that arises from (\ref{Kutact}) 
 it is not completely
clear whether action (\ref{Kutact}) should
be used for the description  of field theory space
around perturbative vacuum $T=0$.  In particular,
in \cite{Berkooz:2002je,Felder:2002sv,
Frolov:2002rr} analysis of the stability of
the  the classical time-dependent
solution of (\ref{tachDBI}) was performed
with some interesting results. It
 was shown that during very short time interval
these fluctuations become large and hence 
could significantly change the classical evolution.
One can also interpret the huge grow of fluctuations
as a creation of the large number of fluctuation
modes in the time-dependent background of
the classical solution \cite{Kluson:2003rd}. 
 Then we can expect that at some
time close to the beginning of the tachyon
condensation the density of the number
of particles created  reaches the stringy density
and the linearised approximation, where 
we presume that the fluctuations are small
breaks down. In fact, the  main goal of this
paper is confirmation of this claim.

In order to calculate 
the number of particle
created at any time event $t$ 
we will formulate the quantum field theory of
fluctuation modes in the Schr\"odinger picture.
Since we will be  interested in the time 
intervals very close to the 
beginning of the rolling tachyon process we will calculate
fluctuation  effective actions in  the regime where
the classical tachyon solution is small. 
We also restrict our analysis to 
the fluctuation modes that have 
 initial momentum $k^2>\mu^2$ at far past.
Modes with $k^2<\mu^2$  exponentially
grow even at the beginning of the tachyon condensation and
significantly change the classical evolution.
The study of the dynamics of these modes is very difficult task
\cite{Felder:2001kt,Felder:2000hj}
and it is certainly beyond the scope of this paper.
We hope that results obtained in this paper
allow us to return to this important problem
in near future.

 Even if we restrict ourselves to the  modes with
the positive frequency $\omk^2=k^2-\mu^2$ at far past we
will get some interesting results. Namely 
we will find  the quantum
field theory of fluctuations that is derived from
the expansion around the classical solution 
of the effective action (\ref{Kutact}) has some
unfamiliar properties.  
 In particular
we will show that for  modes with $k^2>\mu^2$ after short
time interval their time-dependent
 frequencies become imaginary. This fact implies 
huge growth of the number of particles produced
and we can expect the strong backreaction of these
modes on the classical solution.
This result is consistent with the conclusions
presented in \cite{Berkooz:2002je,Felder:2002sv,
Frolov:2002rr}. On the other hand 
in the quantum field theory of fluctuation modes
based on the  tachyon effective
action (\ref{Tperturb}) we will not 
observe such a behaviour. We mean
that this is an indication, of course
not proof, that the effective
action (\ref{Tperturb}) may be considered
as the appropriate one for description of the
tachyon dynamics for small $T$.

This paper is organised as follows. In the next section
(\ref{second}) we will study the quantum field theory
of fluctuation modes around the rolling tachyon 
solution 
on non-BPS D-brane that is described by effective action
(\ref{Kutact}). In section (\ref{third}) we perform the
same analysis in case of bosonic D-brane. In section
(\ref{fourth}) we will study the particle production
on tachyon effective action (\ref{Tperturb}). 
In conclusion (\ref{fifth}) we
outline our results and suggest further directions
of research. 
Finally in Appendix (\ref{sixth}) we give a brief
review of Schr\"odinger picture description of 
quantum field theory.
\section{Non-BPS D-brane effective action}\label{second}
In this section we will study the fluctuation modes
around the rolling tachyon solution on  non-BPS
D-brane in superstring theory. We presume that
the effective field theory description
of the  time-dependent tachyon condensation
 is described by the action (\ref{Kutact}).
Now  the equation of  motion that arises from
(\ref{Kutact}) is
\begin{equation}\label{eqKut}
-\frac{T}{\left(1+\frac{T^2}{2}
\right)^2}\sqrt{\bb}+\frac{T}{2(1+
\frac{T^2}{2})\sqrt{\bb}}-
\partial_{\mu}\left(\frac{\eta^{\mu\nu}
\partial_{\nu}T}{(1+\frac{T^2}{2})\sqrt{\bb}}\right)=0 \ ,
\end{equation}
where
\begin{equation}
\bb=1+\frac{T^2}{2}+\eta^{\mu\nu}\partial_{\mu}T
\partial_{\nu}T \ .
\end{equation}
It is easy to see that
the solution of 
  (\ref{eqKut}) that describes 
the rolling of the tachyon 
from an unstable minimum $T=0$ at $t=-\infty$ to
the stable one $T=\infty$ at $t=\infty$  is  
\begin{equation}\label{solhalf}
T_c(t)=a\sqrt{2}e^{\frac{t}{\sqrt{2}}} \ ,
\end{equation}
where $a$ is some constant which
 can be set to $1$ by time
translation \cite{Lambert:2003zr} .

Now we would like to study 
quantum field theory of
 fluctuation modes above classical solution
(\ref{solhalf}).
 To do this we write the
tachyon field $T$ that appears in
(\ref{Kutact}) as
\begin{equation}\label{Tgen}
T(t,\bx)=T_c(t)+\phi(t,\bx) \ ,
\end{equation}
where $T_c$ is given in (\ref{solhalf}) and
$\phi(t,\bx)$ is fluctuation field around $T_c$
 which is presumed to be small
 with respect to the classical
field  $T_c$. 
 
To get an quadratic
 effective action for fluctuation $\phi$
we insert  (\ref{Tgen}) into (\ref{Kutact})
and perform an expansion with respect to $\phi$
up to the second order. Thanks to the time
dependence of $T_c$ we generally obtain 
an  action for free massive scalar field $\phi$
 where  metric and mass term are time-dependent.
 More precisely,
let us write 
\begin{eqnarray}\label{actexp}
S=-\int dt d^p\bx L(T_c+\phi)=-\int dt d^p\bx
 L(T_c)-\nonumber \\
-\int dt d^p\bx \left[
\frac{\delta L(T_c)}{\delta T}\phi
+\frac{\delta L(T_c)}{\delta \partial_{\mu}T}
\partial_{\mu}\phi\right]-\nonumber \\
-\frac{1}{2}
\int dt d^p\bx \left(\frac{\delta^2 L(T_c)}{
\delta T\delta T}\phi^2+2
\frac{\delta^2 L(T_c)}{
\delta T\delta \partial_{\mu}T}\phi\partial_{\mu}\phi
+\frac{\delta^2 L(T_c)}{
\delta \partial_{\mu}T\delta \partial_{\nu}T}
\partial_{\mu}\phi\partial_{\nu}\phi\right)+\dots \ ,
\nonumber \\ 
\end{eqnarray}
where dots mean terms of higher order in $\phi$. 
Using  equation of motion and integration
by parts we can easily show
that the expression on the
second line in (\ref{actexp}) is equal to zero. 
The quadratic  effective
action for $\phi$  is then given on the third line
in   (\ref{actexp})
\begin{eqnarray}\label{fluctlin}
S=-\frac{1}{2}
\int dtd^p\bx \left[G^{\mu\nu}(T_c)\partial_{\mu}\phi
\partial_{\nu}\phi+M^2(T_c)\phi^2\right] \ , \nonumber \\
G^{\mu\nu}(T_c)=
\frac{\delta^2 L(T_c)}{
\delta \partial_{\mu}T\delta \partial_{\nu}T} \ , \ 
M^2(T_c)=\frac{\delta^2 L(T_c)}
{\delta T \delta T}-\partial_{\mu}\left[
\frac{\delta^2 L(T_c)}{\delta T\delta \partial_{\mu}T}\right] \ .
\nonumber \\
\end{eqnarray}
Let us apply this general description
for the tachyon 
effective action (\ref{Kutact}) and
the solution  (\ref{solhalf}).
After some straightforward calculations
we obtain following values 
of  metric components $G^{\mu\nu}(t)$  and 
mass term $M^2(t)$ that 
appear in (\ref{fluctlin})
\begin{eqnarray}\label{met}
G^{00}=-\frac{V^2}{\sqrt{\bb}}
-\frac{V^2\dot{T}^2}
{
(\bb)^{3/2}}=-1  
\ , 
\nonumber \\
G^{ab}=\frac{V^2\delta^{ab}}{\sqrt{\bb}}=
\frac{\delta^{ab}}{1+e^{\sqrt{2}t}}
\equiv G(t)\delta^{ab} \ ,\nonumber \\
M^2=-\frac{1}{2} \ ,
 \nonumber \\
\end{eqnarray}
where we have used that  for (\ref{solhalf}) 
 $\bb=1+\frac{T^2}{2}-\dot{T}^2=
1$.  As follows  from
(\ref{fluctlin}) and from  (\ref{met})
the quadratic   action for
fluctuation modes has the form of the action
for the scalar field in the time-dependent
background and consequently  we 
can expect that particles will be produced
during the time evolution
\cite{Birrell:ix,Fulling:nb,Wald:yp}.
In order to study this process  
it appears to be useful to formulate 
the quantum field theory of fluctuation
modes in the Schr\"odinger picture description
\footnote{Brief review of Schr\"odinger picture
description of quantum field theory is given
in Appendix.}.  
The main advantage
of the Schr\"odinger picture over other waves
to characterise vacuum states is that it describes
states explicitly by simple wave functional
specified by a single, possibly time-dependent,
kernel function satisfying a differential equation
with a prescribed boundary conditions. This makes
no reference to the assumed spectrum of excited
states and so circumvents the difficulties of the
conventional canonical description of a vacuum
as a no-particle state with respect to the creation
and annihilation operators defined by mode decomposition
of the field, an approach not well suited to
time-dependent problems. In particular, we will
see that using 
explicit time-dependence of the vacuum wave functional
one can easily calculate 
time-dependent vacuum expectation values 
of the quantum field theory operators. This
approach allows us to see at what time the
vacuum expectation value of the operator of the number
of particles with momentum $\bk$ reaches the
string density which has strong consequence
for further evolution of the system. 

Since the brief review of  Schr\"odinger picture description
of quantum field theory is given in the appendix now
we only mention some basic
facts which will be needed in future.  Firstly, 
the wave functional $\Psi$ is function of $\phi(\bx)$ and
explicitly depends on time. Any
operator $\hat{O}(\hphi,\hpi)$ is written
in Schr\"odinger representation as
$\hat{O}(\hphi,\hpi)=
O\left(-i\frac{\delta }{\delta \phi} \ ,
\phi\right)$. Consequently the Schr\"odinger
equation is
\begin{equation}\label{schro}
i\frac{\partial \Psi[\phi,t]}
{\partial t}=H\left(-i\frac{\delta }{
\delta \phi},\phi\right)\Psi[\phi,t] \ .
\end{equation}
Since in the problems we are considering 
the spatial sections are flat, it is natural
to perform a Fourier transformation as
\begin{equation}
\phi(\bx)=\int \ik e^{i\bk \bx}\alpha(\bk) \ . 
\end{equation}
In $k$ space, the Hamiltonian operator that corresponds
to the Lagrangian for fluctuation modes
(\ref{fluctlin}) and that appears in
(\ref{schro}) 
 has the form
\begin{equation}\label{hamki}
H=\frac{1}{2}\int
\ik \left[-\frac{\delta^2}{\delta \alpha(\bk)
\delta\alpha(-\bk)}+
\Omega^2_{\bk}(t)\alpha(\bk)\alpha(-\bk)\right] 
\ ,
\end{equation}
where
\begin{equation}
\Omega^2_{\bk}(t)=\frac{k^2}{1+e^{\sqrt{2}t}}+
m^2 \ , m^2=-\frac{1}{2} \ . 
\end{equation}
We see that for each $k$, the integrand
in (\ref{hamki}) represents
a harmonic oscillator with the 
time-dependent frequency $\Omega^2_{\bk}(t)$. 
Then 
the vacuum wave functional is 
\begin{equation}\label{vacwave}
\Psi[\alpha,t]=
N_0(t)\exp\left[
-\frac{1}{2}\int \ik\alpha(-\bk)
\tilde{G}_{\bk}(t)\alpha(\bk)\right] \ ,
\end{equation}
where the kernel $\tilde{G}_{\bk}(t)$ is solution
of the equation
\begin{equation}
i\frac{\partial \tilde{G}_{\bk}}{\partial
t}=\tilde{G}^2_{\bk}(t)-\Omega^2_{\bk}(t) \ .
\end{equation}
This equation is solved with the ansatz
for $\tilde{G}_{\bk}$ 
\begin{equation}
\tilde{G}_{\bk}(t)=-i\frac{\dot{\psi}_{\bk}(t)}
{\psi_{\bk}(t)} \ , 
\end{equation}
where $\psi_{\bk}$ obeys
\begin{equation}\label{ddpsi}
\ddot{\psi}_{\bk}+\Omega^2_{\bk}(t)\psi_{\bk}=0 \ .
\end{equation}
As we show in the appendix the
vacuum expectation value of the operator 
of the number of  particles with
the momentum $\bk$ at time $t$   is equal to
\begin{eqnarray}
\left<N(\bk,t)\right>=
(2\pi)^{p}\delta_{\bk}(0)
\frac{\left(\Omega_{\bk}(t)-
\tilde{G}_{\bk}(t)\right)\left(
\Omega_{\bk}(t)-\tilde{G}^*_{\bk}(t)\right)}
{2\Omega_{\bk}(t)\left(
\tilde{G}_{\bk}(t)+\tilde{G}_{\bk}^*(t)\right)}
 \ . \nonumber \\
\end{eqnarray}
In the previous expression the
factor $(2\pi)^{p}\delta_{\bk}(0)$ is the volume
of the spatial section hence it is natural to
introduce the vacuum expectation value of
the density of particles with the momentum
$\bk$ $\mathcal{N}(\bk,t)$ as
\begin{eqnarray}\label{vacden}
\left<\mathcal{N}(\bk,t)\right>\equiv
\frac{\left<N(\bk,t)\right>}{V_p}=
\frac{\left(\Omega_{\bk}(t)-
\tilde{G}_{\bk}(t)\right)\left(
\Omega_{\bk}(t)-\tilde{G}^*_{\bk}(t)\right)}
{2\Omega_{\bk}(t)\left(
\tilde{G}_{\bk}(t)+\tilde{G}_{\bk}^*(t)\right)}
 \ . \nonumber \\
\end{eqnarray}
Now it is clear that in order to calculate
(\ref{vacden}) at any time $t$ we should
find the solution of the equation 
(\ref{ddpsi}) that determines the kernel
$\tilde{G}_{\bk}$. 
Since our goal is to determine the vacuum
expectation value of the operator of
the number of particles in time event  near
 the beginning of
the tachyon condensation at $t=-\infty$ 
we have 
$e^{\sqrt{2}t}\ll 1$ and hence
\begin{equation}
\Omega_{\bk}^2(t)=k^2-k^2e^{\sqrt{2}t}+m^2 \ .
\end{equation}
Then (\ref{ddpsi}) has the form
\begin{equation}\label{psi}
\ddot{\psi}_{\bk}-k^2e^{\sqrt{2}t}\psi_{\bk}
+(k^2+m^2)\psi_{\bk}=0
\end{equation}
which can be written as 
\begin{equation}
s^2\psi_{\bk}''+s\psi_{\bk}'-
(s^2+n^2)\psi_{\bk}=0 \ , n^2=-2\omk^2 \ , 
\end{equation}
where
\begin{equation}
s=2\sqrt{\lambda}
e^{\frac{t}{\sqrt{2}}}
\ , 
\lambda=\frac{k^2}{2} \ , \omk^2=k^2+m^2 \ .  
\end{equation} 
The function $\psi_{\bk}$ is the solution
of the second-order ordinary differential
equation and is therefore a linear combination
of two independent solutions with arbitrary
coefficients. Since the kernel itself is 
the derivative of the logarithm $\psi_{\bk}(t)$
the overall normalisation is unimportant. The
kernel therefore depends on one arbitrary parameter
which can be found from the boundary conditions.
Since the initial configuration
corresponds to ordinary unstable D-brane 
it is natural to choose it that
at far past 
the vacuum wave functional
approaches  the usual positive frequency
Minkowski vacuum 
\begin{equation}
\Psi[\phi]\sim \exp \left[-\frac{1}{2}
\int \ik \alpha(\bk)\omega_{0\bk}\alpha(-\bk)
\right] \ . 
\end{equation}
This boundary condition implies that
$\psi_{\bk}$ is 
\begin{equation}
\psi_{\bk}(t)=C_{\bk}J_{\sqrt{2}i\omk}(2\sqrt{\lambda}i
e^{\frac{t}{\sqrt{2}}}) \ .
\end{equation}
To see this note that
for $t\rightarrow -\infty$ 
 $\psi_{\bk}$ has asymptotic form
\begin{equation}
\psi_{\bk}\approx e^{i\omega_{0\bk}t}
\end{equation}
as follows from the
definition of Bessel's functions
\begin{equation}\label{bessel}
J_n(x)=
\left(\frac{x}{2}\right)^n
\sum_{k=0}^{\infty}
\frac{(-1)^k}{k!\Gamma(n+k+1)}
\left(\frac{x}{2}\right)^{2k} \  .
\end{equation}
Then it is easy to see that 
\begin{equation}
\tilde{G}_{\bk}(-\infty)=
\lim_{t\rightarrow -\infty}\left(
-i\frac{\dot{\psi}_{\bk}(t)}{\psi_{\bk}(t)}
\right)
=\omk \ . 
\end{equation}
We use these results in the calculation
of the vacuum expectation value 
(\ref{vacden}) at the time near the beginning 
of the tachyon condensation. For such a large
negative $t$  we can restrict to the terms of 
the second order in  (\ref{bessel}) and we
get
\begin{equation}
\tilde{G}_{\bk}=-i\frac{\dot{\psi}_{\bk}}{\psi_{\bk}}=
 \omk +\mathcal{D}_{\bk}e^{\sqrt{2}t} \ , 
\Omega^2_{\bk}(t)=\omk-\frac{\lambda e^{\sqrt{2}t}}{
\omk}  \ , 
\end{equation}
where
\begin{equation}
\mathcal{D}_{\bk}=
\frac{\left(i(\sqrt{2}-1)-\omk\right)\lambda}
{1+\sqrt{2}i\omk} \ .
\end{equation}
Then the vacuum expectation value 
of the operator of the 
density of particles with momentum  $\bk$
is equal to
\begin{equation}\label{Nestim}
\left<\mathcal{N}(\bk,t)\right>\approx
\frac{1}{2\omk^2}\left(\mathcal{D}_{\bk}+\frac{\lambda}
{\omk}\right)\left(\mathcal{D}_{\bk}^*+\frac{\lambda}
{\omk}\right)e^{2\sqrt{2}t} \ .
\end{equation}
However 
we must stress one important point.
As we claimed  in the introduction we restrict
ourselves to the modes that have positive initial 
frequency at far past $\omk^2=k^2-\frac{1}{2}>0$
that, according to  (\ref{fluctlin})  have 
time-dependent frequencies that for small $t$ are
equal to
\begin{equation}
\Omega^2_{\bk}=\omk^2-
k^2 e^{\sqrt{2}t} \ .
\end{equation}
From the upper expression we can 
deduce that  there exists time event
$t_*$ when $\Omega_{\bk}$ approaches 
zero  
\begin{equation}
\frac{k^2 e^{\sqrt{2}t_*}}{\omk^2}\approx
1 \Rightarrow
e^{\sqrt{2}t_*}\approx \left(1-\frac{1}{2 k^2}\right) \ . 
\end{equation}
We see that for  $k^2\approx \frac{1}{2}$ \ , 
$t_* \rightarrow -\infty$. The vanishing
of $\Omega_{\bk}$ at $t_*$ implies that
 the vacuum
expectation value $\left<\mathcal{N}(\bk,t)\right>$
diverges at $t_*$. This  huge
growth of the vacuum expectation value
of $\mathcal{N}(\bk,t)$ for  $t\rightarrow
t_*$ implies  the
large  backreaction 
on the classical solution 
and linearised approximation breaks down
very quickly after beginning of the tachyon
condensation. 
On the other hand as we will see in section 
(\ref{fourth}) such a behaviour is
absent in the case of tachyon effective
action (\ref{Tperturb}).  

In spite of this remark   
one can use  (\ref{Nestim})
for the estimation of 
 the time when the density of particles
with $k^2\gg 1/2 $ is of order of the string
scale. In this case
we have 
$\omk \approx k \ , \frac{\lambda}{\omk}
\approx k \ 
 \ , \mathcal{D}_{\bk}
\approx i k^2 $  and hence
\begin{equation}
\left<\mathcal{N}(\bk,t)\right>\approx 
\frac{1}{k^2}(k+ik^2)(k-ik^2)e^{2\sqrt{2}t}\approx
\frac{(k^2+k^4)}{k^2}e^{2\sqrt{2}t}
\approx k^2e^{2\sqrt{2}t}
\end{equation}
so that the time when the 
$\left<\mathcal{N}(\bk,t)\right>$
reaches stringy density is
proportional to
\begin{equation}
\left<\mathcal{N}(\bk,t)\right>\approx 1 
=k^2e^{2\sqrt{2}t_s}
\Rightarrow 
e^{\sqrt{2}t_s}\approx \frac{1}{k}\ll 1 \ ,
t_s\approx -\ln k \ll 0 \  
\end{equation}
that is again close to the beginning
of the tachyon condensation. Note 
that  for   $t=t_s$ 
 $\Omega_{\bk}(t_s)$ is still 
positive 
\begin{equation}
\Omega^2_{\bk}(t_s)=
\omk^2-k^2e^{\sqrt{2}t_s}\approx
k(k-1)\gg 1 \ .
\end{equation}
To conclude,  the fact that the number of
particles that are created during very short
 time evolution of the 
vacuum state is so large, leads to the 
breaking of the the linearised approximation 
 very quickly after
beginning of the tachyon condensation. 
Then we can also expect  large
backreaction of fluctuation modes on the classical
solution.   This result is 
in agreement with analysis of fluctuations
performed in 
\cite{Berkooz:2002je,Felder:2002sv} where
several  tachyon ``DBI'' actions were studied.
\section{Particle production on 
D-brane in bosonic theory}\label{third}
In this section we will calculate the particle production
on  D-brane in bosonic string theory
when we presume the tachyon effective
action has the form  
\footnote{For recent discussion of the effective
D-brane action in bosonic theory, see
\cite{Smedback:2003ur}.}
\begin{equation}\label{Tbos}
S=-\int dt d^p\bx 
\frac{1}{1+T}\sqrt{1+T+\frac{\eta^{\mu\nu}
\partial_{\mu}T\partial_{\nu}T}{T}} \ .
\end{equation}
This  tachyon effective action 
was derived in \cite{Kluson:2003sr} according
to the proposal given  in 
\cite{Kutasov:2003er}. Even if there are
some problems with this procedure thanks to
the non-analycity around the point $T=0$
\cite{Kutasov:2003er} one can   consider
this action as another model describing
the tachyon dynamics on Dp-brane in the bosonic theory
around the conformal point $T\approx e^t$. In
fact, the  equation of motion
that arises from (\ref{Tbos})
 is
\begin{equation}\label{boseq}
-\frac{1}{(1+T)^2}
\sqrt{\bb}+\frac{1}
{2(1+T)\sqrt{\bb}}-
\frac{\eta^{\mu\nu}\partial_{\mu}T
\partial_{\nu}T}{2(1+T)T^2\sqrt{\bb}}
-
\partial_{\mu}\left(\frac{1}{1+T}
\frac{\eta^{\mu\nu}\partial_{\nu}T}{T
\sqrt{\bb}}\right)=0 \ . 
\end{equation}
Then one can show that  $T_c=e^t$
is solution of the equation of motion (\ref{boseq}).
Since we
will be mainly interested in the particle production
shortly after the beginning of the rolling tachyon 
process we restrict ourselves to the case when
$T, \frac{\eta^{\mu\nu}\partial_{\mu}T
\partial_{\nu}T}{T}$ are small.
Then we get the tachyon effective action
up to the second order
of $T$
\begin{eqnarray}\label{Tsecond}
S=-\int dt d^p\bx \left(
\frac{\eta^{\mu\nu}\partial_{\mu}T
\partial_{\nu}T}{2T}-\frac{T}{2}
+\frac{3T^2}{8}
-\frac{3\eta^{\mu\nu}\partial_{\mu}T
\partial_{\nu}T}{4}-
\frac{(\eta^{\mu\nu}\partial_{\mu}T
\partial_{\nu}T)^2}{8T^2}\right)=
\nonumber \\
=-\int dt d^p\bx \left(
2\eta^{\mu\nu}
\partial_{\mu}t\partial_{\nu}t-\frac{t^2}{2}
+\frac{3t^4}{8}-3t^2
\eta^{\mu\nu}
\partial_{\mu}t\partial_{\nu}t
-2(\eta^{\mu\nu}
\partial_{\mu}t\partial_{\nu}t)^2
\right) \ , \nonumber \\
\end{eqnarray}
where in the second line we have performed
the field redefinition
\begin{equation}
T=t^2 \ , \partial_{\mu}T=
2t\partial_{\mu}t \ .
\end{equation}
The equation of motion that arises
from (\ref{Tsecond}) is
\begin{eqnarray}
-4\partial_{\mu}\left(\eta^{\mu\nu}
\partial_{\nu}t\right)-t+\frac{3
t^3}{2}
-6t\eta^{\mu\nu}
\partial_{\mu}t\partial_{\nu}t+
6\partial_{\mu}\left(t^2\eta^{\mu\nu}
\partial_{\nu}t\right)+\nonumber \\
+
8\partial_{\mu}\left[
(\eta^{\kappa\gamma}\partial_{\kappa}t
\partial_{\gamma}t)\eta^{\mu\nu}
\partial_{\nu}t\right]=0 \ . \nonumber \\
\end{eqnarray}
One can show that this equation of motion
has an exact solution  $t_c=e^{\beta t}$ 
with $\beta=1/2$. 

Now we will study the fluctuation modes
around the classical solution $t_c$. 
After performing the same calculations
as in the previous section we get
the metric and the mass term in
(\ref{fluctlin})
\begin{eqnarray}
G^{00}=\frac{\delta^2 L}{\delta \partial_0t
\delta \partial_0 t}=-4 \ ,
G^{ab}=\frac{\delta^2 L}{\delta \partial_{a}t
\delta \partial_{b}t}=4\delta^{ab}(1-e^{t}) 
\nonumber \\
M^2=\frac{\delta^2 L}{\delta^2t} 
-\partial_{\mu}\left(
\frac{\delta^2L}{\delta t\delta
\partial_{\mu}t}\right)=-1 \ 
\nonumber \\
\end{eqnarray}
and hence  the quadratic action
 for fluctuation modes is
\begin{eqnarray}
S=-\frac{1}{2}\int dt d^p\bx
\left[-4\partial_t \phi \partial_t \phi+
4\delta^{ab}(1-e^{t})\partial_a \phi
\partial_b \phi-\phi^2\right]
\Rightarrow \nonumber \\
S=-\frac{1}{2}\int dt d^p\bx
\left[-\partial_t \psi \partial_t \psi+
\delta^{ab}(1-e^{t})\partial_a \psi
\partial_b \psi-\frac{1}{4}\psi^2\right]
 \nonumber \\
\end{eqnarray}
after rescaling $2\phi=\psi$. 
Now we will proceed as in the previous section.
Firstly, the Hamiltonian operator has 
the form
\begin{equation}
H=\frac{1}{2}
\int \ik\left[-\frac{\delta^2}{\delta
\alpha(-\bk)\delta\alpha(\bk)}+\Omega^2_{\bk}(t)
\alpha(-\bk)\alpha(\bk)\right] \ ,  
\end{equation}
where
\begin{equation}\label{omegabos}
\Omega^2_{\bk}(t)=(1-e^t)k^2
-\frac{1}{4}
=\omk^2-\lambda e^t \ , 
\omk^2=k^2-\frac{1}{4} \ ,
\lambda=k^2 \ . 
\end{equation}
We see that the only difference with the
analysis given in previous section is in
the form of  (\ref{omegabos}). So that
after performing 
the same calculation as in section
(\ref{second})  we obtain that 
 kernel $\tilde{G}_{\bk}$ in 
(\ref{vacwave}) is equal to
\begin{equation}
\tilde{G}_{\bk}(t)=-i\frac{\dot{f}_{\bk}}
{f_{\bk}} 
\ ,
f_{\bk}=
CJ_{2i\omk}(2\sqrt{\lambda}e^{t/2}) \ 
\end{equation}
and consequently 
the vacuum expectation value of the
operator of density of particles with 
momentum $\bk$ at time close to the
beginning of the tachyon condensation
is equal to
\begin{equation}\label{denbos}
\left<\mathcal{N}(\bk,t)\right>\approx
\frac{1}{2\omk^2}\left(\omk\mathcal{D}_{\bk}+\frac{\lambda}
{\omk}\right)\left(\omk\mathcal{D}_{\bk}^*+\frac{\lambda}
{\omk}\right)e^{2t} \ , 
\end{equation}
where
\begin{equation}
\mathcal{D}_{\bk}=
\frac{\lambda}{(1+i\omk)} \ . 
\end{equation}
From (\ref{omegabos}) we   see that modes with
$k\approx 1/2$ have time-dependent frequencies
$\Omega_{\bk}$ that approach zero very quickly 
after beginning of the rolling tachyon process
\begin{equation}
\omk^2-k^2e^{t_*}\approx 0
\Rightarrow e^{t_*}=
1-\frac{1}{4k^2}  \ .
\end{equation}
The interpretation of this problem is the same
as in the Supersymmetric case discussed in
previous section. 
On the other hand for modes with $k\gg 1/2$ we have
$ \omk\mathcal{D}_{\bk}\approx -ik^2 \ ,
\frac{\lambda}{\omk}\approx k$ and hence
(\ref{denbos}) is
\begin{equation}
\left<\mathcal{N}(\bk,t)\right>\approx\frac{1}{k^2}
(k-ik^2)(k+ik)^2e^{2t}\approx k^2e^{2t}
\end{equation}
so that the time $t_s$ when the vacuum
expectation value of density of particles reaches 
the stringy density is
\begin{equation}
\left<\mathcal{N}(\bk,t_s)\right>\approx 1
\Rightarrow t_s\approx -\ln k 
\end{equation}
which for large $k$ again implies that the 
density of particles with momentum $\bk \gg 1$ 
become stringy very shortly after beginning
of the tachyon condensation.   
\section{Perturbative tachyon effective action}
\label{fourth}
In this section we will perform the analysis
of fluctuations around the  rolling
tachyon solution of the equation of motion
that arises from (\ref{Tperturb}) 
\begin{equation}\label{Tperturb1}
S=-\frac{1}{2}\int dtd^p\bx \left[
\eta^{\mu\nu}\partial_{\mu}T
\partial_{\nu}T-\mu^2T^2+g T^4+
\dots\right] \ , 
\end{equation}
where $\mu^2_{bose}=1 \ ,
\mu^2_{super}=\frac{1}{2}$. 
The linearised  equation of motion that follows
from (\ref{Tperturb1}) is
\begin{equation}
\partial_{\mu}\left(
\eta^{\mu\nu}\partial_{\nu}T\right)+
\mu^2T=0 \ 
\end{equation}
that has  time-dependent solution 
\begin{equation}\label{Tc}
T_c=e^{\lambda t}  \ ,   
 -\lambda^2+\mu^2=0  \ .
\end{equation}
Note that one can trust
the linearised approximation  for
the classical field $T_c$ only if 
  $\mu^2T^2\gg gT^4
\Rightarrow \frac{\mu^2}{g}\gg e^{2t}$ and hence
we must restrict to the times near the
begining of the tachyon condensation when
 $e^{2t}\ll 1$. 

In order to study the dynamics of 
fluctuations around (\ref{Tc}) we will proceed
as in  section (\ref{second}). From 
(\ref{Tperturb1}) we get the metric and
mass term in the quadratic  action
for fluctuation field $\phi$
(\ref{fluctlin})
\begin{eqnarray}
G^{\mu\nu}=\frac{\delta^2 L}{\delta \partial_{\mu}T
\partial_{\nu}T}=-\eta^{\mu\nu} \ ,
\nonumber \\
m^2(t)=\frac{\delta^2 L}{\delta T\delta T}=
\mu^2-6gT^2_{c}=
\mu^2-6ge^{2\mu t} \nonumber \\
\end{eqnarray}
So that the Lagrangian and Hamiltonian $H$ are
\begin{eqnarray}
L=-\frac{1}{2}\int dt d^p\bx \left[
\eta^{\mu\nu}\partial_{\mu}\phi
\partial_{\nu}\phi-(\mu^2-6ge^{2\mu t})\phi^2\right] 
\Rightarrow 
\nonumber \\
H=\frac{1}{2}\int d^p\bx\left[\dot{\phi}^2+
\delta^{ab}\partial_a\phi
\partial_b\phi+(6ge^{2\mu t}-\mu^2)\phi^2
\right] \ . \nonumber \\
\end{eqnarray}
Then  the corresponding operator $\hat{H}$ is
\begin{eqnarray}\label{Hoper}
\hat{H}=\frac{1}{2}\int \ik\left[-\frac{\delta^2}
{\delta \alpha(\bk)\alpha(-\bk)}+\Omega^2_{\bk}(t)
\alpha(-\bk)\alpha(\bk)\right]
\nonumber \\
\Omega^2_{\bk}(t)=
6ge^{2\mu t}+\omk^2\equiv
\lambda e^{2\mu t}+\omk^2
 \ ,
\omk^2=k^2-\mu^2 \ , \lambda=6g \ . 
\end{eqnarray}
In the following we restrict ourselves
to the bosonic case $\mu^2=1$, for supersymetric
case $\mu^2=1/2$ the situation will be 
completely the same. 
As in previous section we are interested
in the form of the kernel $\tilde{G}_{\bk}(t)=
-i\frac{\dot{f}_{\bk}}{f_{\bk}}$ where
$f_{\bk}$ obeys the differential equation
\begin{equation}\label{fker}
\ddot{f}_{\bk}+\Omega^2_{\bk}f_{\bk}=0 \ .
\end{equation}
For $\Omega$ given in (\ref{Hoper}) the
equation (\ref{fker})  has the general
solution
\begin{equation}\label{fexp}
f_{\bk}(t)=Af^{in}_{\bk}(t)+
B f^{in *}_{\bk}(t) \ ,
\end{equation}
where
\begin{equation}
f_{\bk}^{in}(t)=
J_{-i\omk}
(\sqrt{\lambda}e^{t}) \ . 
\end{equation}
The  condition  that the
vacuum wave functional approaches for $t\rightarrow -\infty$ 
the  usual Minkowski vacuum state functional
implies that  $A=0, B=1$ in (\ref{fexp})
so that
\begin{equation}
\tilde{G}_{\bk}(t)=-i\frac{\dot{
f}_{\bk}^{in*}}{f_{\bk}^{in*}} \ . 
\end{equation}
Repeating the same calculations as
in section (\ref{second}) we obtain 
the vacuum expectation value of the operator
of the density of particles with momentum
$\bk$ at time close to the beginning
of the tachyon condensation 
\begin{equation}
\left<\mathcal{N}(\bk,t)\right>\approx
\frac{1}{2}\left(
\frac{\lambda}{2\omk^2}+D_{\bk}\right)
\left(\frac{\lambda}{2\omk^2}+D^*_{\bk}\right)
e^{4t} \ , 
\end{equation}
where
\begin{equation}
D=\frac{\lambda}{(1+i\omk)} \ .
\end{equation}
For large $\omk \gg 1$ the upper expression simplifies
to 
\begin{equation}\label{largeomega}
\left<\mathcal{N}(\bk,t)\right>\approx
\frac{\lambda^2}{\omk^4}e^{4t} \ . 
\end{equation}
Since  the linearised approximation 
breaks down when $\mu^2 T^2\approx g T^4 \Rightarrow
e^{2t_{lin}}=\frac{6}{\lambda}$ we obtain from
(\ref{largeomega}) that for  $t=t_{lin}$ the number
of particles produced remains still small
\begin{equation}
\left<\mathcal{N}(\bk,t_{lin})\right>\approx
\frac{1}{\omk^4}\ll 1 \ . 
\end{equation}
We also see that as opposite to
the cases of the effective actions discussed
in previous sections the  time-dependent frequency
$\Omega_{\bk}^2$ is  positive for 
all times  so that   the vacuum expectation
value of the operator of density of particles
do not diverge,
or equivalently 
\cite{Berkooz:2002je,Felder:2002sv,Frolov:2002rr}
  no new exponentially
growing modes are produced during the time evolution. 
From these facts  one can
deduce  that the effective action
(\ref{Tperturb}) correctly describes  the tachyon
dynamics at the beginning of the tachyon condensation. 
\section{Conclusion}\label{fifth}
In this paper we have calculated the production
of fluctuation modes around the rolling tachyon
solution on unstable D-brane at times near 
the beginning of the tachyon condensation. 
The main purpose of this analysis was to see 
whether creation of fluctuation modes in the 
time-dependent background of the classical
solution is
strong enough to be able to reach the stringy
density at finite time from the beginning
of the rolling tachyon process. Then
  we can expect that the
linearised approximation  breaks down 
 and in order to properly describe the
tachyon condensation on D-brane 
 we should study the full non-linear action
and also take  into
account the  backreaction of fluctuations
on classical solution. Unfortunately these
problems are very difficult and their
solution is beyond the scope of this paper. 
We also mean that the calculation of 
the particle production could be useful in the
answering the question which form of the
tachyon effective action is more suitable for
the description of the beginning of the
tachyon condensation.
 In particular, 
this question seems to be very important in case
of effective action (\ref{Kutact}) since its
precise meaning is not completely clear. 
 Recently, very nice
analysis of the tachyon effective action
(\ref{Kutact}) was performed where it
was argued that there are 
problems with the interpretation of
(\ref{Kutact}) or more generally with
action  (\ref{tachDBI}).
On the other hand  it was shown in the recent paper 
\cite{Sen:2003zf} that (\ref{tachDBI})
gives  description of the moduli
space of D-brane compactified on 
circle of critical radius which is
in very good agreement with the exact
CFT analysis. 

The conclusion given above is directly
related to the results we have found
in this paper when we have
studied the quantum  field theory of fluctuation modes
around the rolling tachyon solution
of the equation of motion that arises
from   (\ref{Kutact}).
In particular, we have shown that for modes, that initially
correspond to ordinary fluctuation modes with
positive frequencies, their effective
time-dependent frequencies vanish  
at time  near  the beginning of
the tachyon condensation and consequently
 the vacuum expectation value of
the particle density diverges 
 very shortly after beginning of the 
tachyon condensation. 
We mean that this result is in 
agreement with the analysis performed in
 \cite{Berkooz:2002je,Felder:2002sv,
Frolov:2002rr} where  it was shown instability of
the classical solution of the equation motion
arising from (\ref{tachDBI}) 
 with respect to the growth of
the fluctuation modes. It was argued there that 
the linear approximation breaks down very quickly after
beginning of the rolling tachyon process.
On the other hand when we have performed the same analysis
in case of the tachyon effective action 
(\ref{Tperturb})  we 
have found that the particle productions of 
positive frequency modes is small in the 
time interval when we can trust the linearised
approximation. These results suggest to us that
 the effective action
(\ref{Tperturb}) seems to be suitable for
the  description
of the tachyon dynamics at the beginning of the 
rolling tachyon process. 

As we mentioned above 
to find the  regions of validity of
various tachyon effective actions is an 
important problem. Its solving could
lead to better understanding of the
problem of the tachyon condensation
in string theory. It is clear that there
are lot of open questions 
in the effective action description of
the tachyon dynamics on D-brane and
we also hope that our modest contribution
could be helpful in  further research
in this area.

\section{Appendix: QFT in Schr\"odinger picture}
\label{sixth}
In this appendix we briefly review the Schr\"odinger
picture formulation of quantum field theory.
For more informations, see \cite{Long:1996wf,Guven:1987bx}.

Let us star with the action for free scalar
field with time-dependent mass term 
\footnote{Our convention is
$\eta_{\mu\nu}=\mathrm{diag}(-1,\dots, 1) \ ,
\mu\ , \nu=0,\dots, p \ ,
x^0=t \ , \bx=(x^1,\dots,x^p)$.}
\begin{equation}
S=\int dtd^p\bx L=
-\frac{1}{2}\int dtd^p\bx\left(
-\partial_t\phi\partial_t\phi+\eta^{ab}
\partial_a\phi\partial_b\phi+m^2(t)\phi^2
\right) \ .
\end{equation}
The canonical momentum conjugate to $\phi$ is
\begin{equation}
\Pi(t,\bx)=\frac{\delta L}{\delta \dot{\phi}}
=\dot{\phi} \ ,
\end{equation}
and the Hamiltonian
\begin{equation}
H=\int d^p\bx\left(\Pi \dot{\phi}-L\right)=
\frac{1}{2}\int d^p\bx
\left(\Pi^2+\eta^{ab}\partial_{a}\phi
\partial_b\phi+m^2(t)\phi^2\right) \ .
\end{equation}
The system can be canonically quantised by treating the
fields as operators and imposing appropriate commutation
relations. This involves choice of a foliation
of a space-time into a succession of
spacelike hypersurfaces. We choose these to
be the hypersurface of fixed $t$ and
impose equal time commutation relations 
\begin{equation}
[\hphi(\bx,t),\hpi(\by,t)]=
i\delta (\bx-\by) \ , [\hphi(\bx,t),\hphi(\by,t)]=
[\hpi(\bx,t),\hpi(\by,t)]=0 \ .
\end{equation}
In the Schr\"odinger picture 
we take the basis vector of the state 
vector space to be the eigenstate 
of the field operator $\hphi(t,\bx)$
on a fixed $t$ hypersurface, with eigenvalues $\phi(\bx)$
\begin{equation}
\hphi(t,\bx)\ket{\phi(\bx),t}=
\phi(\bx)\ket{\phi(\bx),t} \ .
\end{equation}
Notice that the set of field eigenvalues $\phi(\bx)$
is independent of the value of $t$ labelling the
hypersurface.  In this picture, the quantum states
are explicit functions of time and are represented
by wave functionals $\Psi[\phi(\bx),t]$. Operators
$\hat{\mathcal{O}}(\hphi,\hpi)$ acting on these
states may be represented by
\begin{equation}
\hat{\mathcal{O}}(\hpi(x),\hphi(x))=
\mathcal{O}\left(-i\frac{\delta }{\delta
\phi(\bx)},  \phi(\bx)\right) \ .
\end{equation}
The Schr\"odinger equation which governs the evolution
of the wave functional between the spacelike 
hypersurfaces, is
\begin{eqnarray}\label{schrod}
i\frac{\partial \Psi[\phi,t]}
{\partial t}=H\left(-i\frac{\delta }{\delta
\phi(\bx)},  \phi(\bx)\right)\Psi[\phi,t]=\nonumber \\
=\frac{1}{2}\int
d^p\bx \left[-\frac{\delta^2}{\delta \phi^2}
+\eta^{ab}\partial_a\phi\partial_b\phi+
m^2(t)\phi^2\right]\Psi[\phi,t] \ . 
\nonumber \\ 
\end{eqnarray}
To solve this equation, we make the ansatz, that up
to time-dependent phase, the vacuum functional
 is simple
Gaussian.  We therefore write
\begin{equation}
\Psi_0[\phi,t]=
N_0(t)\exp\left\{-\frac{1}{2}\int
d^p\bx d^p\by \phi(\bx)G(\bx,\by,t)\phi(\by)\right\}=
N_0(t)\psi_0(\phi,t) \ ,
\end{equation}
where $N_0(t)$ and $G(\bx,\by,t)$ obey
\begin{eqnarray}\label{kerA}
i\frac{\partial N_0(t)}{\partial t}=
\frac{N_0(t)}{2}\int d^p\bz
G(\bz,\bz,t) \ , \nonumber \\
i\frac{\partial G(\bx,\by,t)}
{\partial t}=
\int d^p\bz  G(\bz,\bx,t)G(\by,\bz,t)
-\left(\eta^{ab}\partial_a \partial_b
+m^2(t)\right)\delta(\bx,\by) \ . 
\nonumber \\
\end{eqnarray}
For situations, where
 the spatial sections are flat, 
it is natural
to perform a Fourier transformation on
the space dependence of field configuration
\begin{equation}
\phi(\bx)=\int \ik e^{i\bk \bx}\alpha(\bk)
\end{equation}
Similarly we can define $\delta/\delta \alpha(\bk)$
as 
\begin{equation}
\frac{\delta }{\delta \phi(\bx)}=
\int \ik e^{i\bk \bx}\frac{\delta }{
\delta \alpha(\bk)} \ .
\end{equation}
Reality of $\phi$ implies $\alpha^*(\bk)=
\alpha(-\bk)$ and since
\begin{equation}
\frac{\delta \phi(\bx)}{
\delta \phi(\bx')}=
\delta (\bx-\bx') 
\end{equation}
we also have
\begin{equation}
\frac{\delta \alpha(\bk)}{
\delta \alpha(\bk')}=
(2\pi)^p\delta
(\bk+\bk') \ . 
\end{equation}
In $k$ space, the Hamiltonian is
\begin{equation}\label{hamk}
H=\frac{1}{2}\int
\ik \left[-\frac{\delta^2}{\delta \alpha(\bk)
\delta\alpha(-\bk)}+
\Omega^2_{\bk}(t)\alpha(\bk)\alpha(-\bk)\right] 
\ ,
\end{equation}
where
\begin{equation}
\Omega^2_{\bk}(t)=m^2(t)+\omk^2 \ .
\omk^2=k^2+m^2
\end{equation}
The Fourier transform of the kernel
$G(\bx,by,t)$ is
\begin{equation}
G(\bx,\by,t)=\int \ik
e^{i\bk(\bx-\by)}\tg(\bk,t) \  
\end{equation}
and hence the kernel equation
(\ref{kerA})
reduces to 
\begin{equation}
i\frac{\partial \tg(\bk,t)}
{\partial t}=\tg^2(\bk,t)-
\Omega^2_{\bk}(t) \ ,
\Omega^2_{\bk}(t)=k^2+m^2(t) 
\end{equation}
which can be solved with the
ansatz
\begin{equation}
\tg(\bk,t)=-i\frac{\dot{\psi}_{\bk}(t)}{
\psi_{\bk}(t)} \ ,
\end{equation}
where  $\psi_{\bk}(t)$ obeys
\begin{equation}\label{psieq}
\ddot{\psi}_{\bk}+\Omega^2_{\bk}(t)
\psi_{\bk}=0 
\end{equation}
Note that (\ref{psieq})
is differential equation of second order so that
 its general solution is given as
linear combination of two solutions and
hence is parametrised by two independent constants.
On the other hand $\psi$ appears in $\tg$ only
in combination $\frac{\dot{\psi}_{\bk}}{\psi_{\bk}}
$ and hence overall normalisation constant
is unimportant. As a result the vacuum functional
 depends on  one free parameter. The value
of this parameter can be fixed from the initial
conditions that vacuum wave functional should
obey.  

In order to define concept of operator of 
particles in the quantum field theory with
time-dependent mass we will proceed as follows.
Let us define following set of
operators
\begin{eqnarray}\label{aadag}
\ma(\bk,t)=\frac{1}{\sqrt{2\Omega_{\bk}}}
\left[\frac{\delta }{\delta \alpha(\bk)}+\Omega_{\bk}
(t)\alpha(\bk)\right] \nonumber \\
\ma^{\dag}(\bk,t)=
\frac{1}{\sqrt{2\Omega_{\bk}}}
\left[-\frac{\delta }{\delta \alpha(-\bk)}+\Omega_{\bk}
(t)\alpha(-\bk)\right] \nonumber \\
\ma^{\dag}(\bk,t)\ma(\bk,t)=
\frac{1}{2\Omega_{\bk}}\left[
-\frac{\delta^2}{\delta \alpha(-\bk)
\alpha(\bk)}+\Omega^2_{\bk}(t)\alpha(\bk)
\alpha(-\bk)-\Omega_{\bk}(t)(2\pi)^p
\delta_{\bk}(0)\right] \ . \nonumber \\
\end{eqnarray}
With the help of (\ref{aadag}) we 
can  rewrite the Hamiltonian 
as
\begin{equation}
H(t)=\frac{1}{2}
\int \ik \Omega_{\bk}(t)\left[
\ma^{\dag}(\bk,t)\ma(\bk,t)+V_p\right]
\ , V_p=(2\pi)^p \delta_{\bk}(0) \  
\end{equation}
that has the form of sum of 
an infinite number of Hamiltonians
 of harmonic oscillators with the 
time-dependent frequencies $\Omega_{\bk}(t)$ 
where one can interpret
 $\ma(\bk,t)(\ma^{\dag}(\bk,t))$ 
as operators that annihilate(create) particle
with momentum $\bk$ at time $t$. According
to this interpretation the operator
of the number of particles with momentum $\bk$ 
at time $t$ is 
\begin{equation}
N(\bk,t)=\ma^{\dag}(\bk,t)\ma(\bk,t) \ . 
\end{equation}
We are mainly interested in the vacuum expectation
value of this operator $\left<N_{\bk}(t)\right>$.
It can be shown \cite{Long:1996wf,Guven:1987bx}
\footnote{For recent calculation of this vacuum
expectation value, see \cite{Kluson:2003sh}.}
that its vacuum expectation value 
is equal to
\begin{eqnarray}
\left<N(\bk,t)\right>=
(2\pi)^p\delta_{\bk}(0)
\frac{\left(\Omega_{\bk}(t)-\tg(\bk,t)\right)
\left(\Omega_{\bk}(t)-\tg^*(\bk,t)\right)
}{2\Omega_{\bk}(t)(\tg(\bk,t)+\tg^*(\bk,t))} \ .
 \nonumber \\
\end{eqnarray}
Since the factor $(2\pi)^p\delta_{\bk}(0)=V_p$
is volume of the spatial section of Dp-brane
it is natural to define the vacuum expectation
value of the operator of density of 
particles with momentum $\bk$ at time $t$ 
$\mathcal{N}(\bk,t)$ as
\begin{equation}\label{mathcalN}
\left<\mathcal{N}(\bk,t)\right>
\equiv \frac{\left<N(\bk,t)\right>}{V_p}=
\frac{\left(\Omega_{\bk}(t)-\tg(\bk,t)\right)
\left(\Omega_{\bk}(t)-\tg^*(\bk,t)\right)
}{2\Omega_{\bk}(t)(\tg(\bk,t)+\tg^*(\bk,t))} \ .
\end{equation}
\\
\\
{\bf Acknowledgement}
This work was supported by the
Czech Ministry of Education under Contract No.
14310006.
\\
\\


\begin{thebibliography}{20}
\bibitem{Sen:2002an}
A.~Sen,
\emph{``Field theory of tachyon matter,''}
Mod.\ Phys.\ Lett.\ A {\bf 17} (2002) 1797
[arXiv:hep-th/0204143].


\bibitem{Sen:2002in}
A.~Sen,
\emph{``Tachyon matter,''}
JHEP {\bf 0207} (2002) 065
[arXiv:hep-th/0203265].


\bibitem{Sen:2002nu}
A.~Sen,
\emph{``Rolling tachyon,''}
JHEP {\bf 0204} (2002) 048
[arXiv:hep-th/0203211].

\bibitem{Gutperle:2003xf}
M.~Gutperle and A.~Strominger,
\emph{``Timelike boundary Liouville theory,''}
Phys.\ Rev.\ D {\bf 67} (2003) 126002
[arXiv:hep-th/0301038].










\bibitem{Strominger:2002pc}
A.~Strominger,
\emph{``Open string creation by S-branes,''}
arXiv:hep-th/0209090.


\bibitem{Maloney:2003ck}
A.~Maloney, A.~Strominger and X.~Yin,
\emph{``S-brane thermodynamics,''}
arXiv:hep-th/0302146.

\bibitem{Sen:2003tm}
A.~Sen,
\emph{``Dirac-Born-Infeld action on the tachyon kink and vortex,''}
arXiv:hep-th/0303057.





\bibitem{Sen:2002qa}
A.~Sen,
\emph{``Time and tachyon,''}
arXiv:hep-th/0209122.




\bibitem{Kutasov:2003er}
D.~Kutasov and V.~Niarchos,
\emph{``Tachyon effective actions in open string theory,''}
arXiv:hep-th/0304045.


\bibitem{Lambert:2002hk}
N.~D.~Lambert and I.~Sachs,
\emph{``Tachyon dynamics and the effective action approximation,''}
Phys.\ Rev.\ D {\bf 67} (2003) 026005
[arXiv:hep-th/0208217].

\bibitem{Lambert:2001fa}
N.~D.~Lambert and I.~Sachs,
\emph{``On higher derivative terms in tachyon effective actions,''}
JHEP {\bf 0106} (2001) 060
[arXiv:hep-th/0104218].

\bibitem{Sen:2003bc}
A.~Sen,
\emph{``Open and closed strings from unstable D-branes,''}
arXiv:hep-th/0305011.



\bibitem{Lambert:2003zr}
N.~Lambert, H.~Liu and J.~Maldacena,
\emph{``Closed strings from decaying D-branes,''}
arXiv:hep-th/0303139.


\bibitem{Kim:2003in}
C.~Kim, Y.~Kim and C.~O.~Lee,
\emph{``Tachyon kinks,''}
arXiv:hep-th/0304180.

\bibitem{Garousi:2003ai}
M.~R.~Garousi,
\emph{``Slowly varying tachyon and tachyon potential,''}
arXiv:hep-th/0304145.






\bibitem{Sen:1999md}
A.~Sen,
\emph{``Supersymmetric world-volume action for non-BPS D-branes,''}
JHEP {\bf 9910} (1999) 008
[arXiv:hep-th/9909062].

\bibitem{Garousi:2000tr}
M.~R.~Garousi,
\emph{``Tachyon couplings 
on non-BPS D-branes and Dirac-Born-Infeld action,''}
Nucl.\ Phys.\ B {\bf 584} (2000) 284
[arXiv:hep-th/0003122].

\bibitem{Bergshoeff:2000dq}
E.~A.~Bergshoeff, M.~de Roo, T.~C.~de Wit, E.~Eyras and S.~Panda,
\emph{``T-duality and actions for non-BPS D-branes,''}
JHEP {\bf 0005} (2000) 009
[arXiv:hep-th/0003221].

\bibitem{Kluson:2000iy}
J.~Kluson,
\emph{``Proposal for non-BPS D-brane action,''}
Phys.\ Rev.\ D {\bf 62} (2000) 126003
[arXiv:hep-th/0004106].



\bibitem{Minahan:2000ff}
J.~A.~Minahan and B.~Zwiebach,
\emph{``Field theory models for 
tachyon and gauge field string dynamics,''}
JHEP {\bf 0009} (2000) 029
[arXiv:hep-th/0008231].


\bibitem{Minahan:2000tf}
J.~A.~Minahan and B.~Zwiebach,
\emph{``Effective tachyon dynamics 
in superstring theory,''}
JHEP {\bf 0103} (2001) 038
[arXiv:hep-th/0009246].

\bibitem{Tseytlin:2000mt}
A.~A.~Tseytlin,
\emph{``Sigma model approach to 
string theory effective actions with tachyons,''}
J.\ Math.\ Phys.\  {\bf 42} (2001) 2854
[arXiv:hep-th/0011033].






\bibitem{Kluson:2003sr}
J.~Kluson,
\emph{``Note on D-brane effective action in 
the linear dilaton background,''}
arXiv:hep-th/0310066.

\bibitem{Smedback:2003ur}
M.~Smedback,
\emph{``On effective actions for the bosonic tachyon,''}
arXiv:hep-th/0310138.

\bibitem{Fotopoulos:2003yt}
A.~Fotopoulos and A.~A.~Tseytlin,
\emph{``On open superstring partition 
function in inhomogeneous rolling tachyon background,''}
arXiv:hep-th/0310253.




\bibitem{Felder:2001kt}
G.~N.~Felder, L.~Kofman and A.~D.~Linde,
\emph{``Tachyonic instability and 
dynamics of spontaneous symmetry breaking,''}
Phys.\ Rev.\ D {\bf 64} (2001) 123517
[arXiv:hep-th/0106179].

\bibitem{Felder:2000hj}
G.~N.~Felder, J.~Garcia-Bellido, P.~B.~Greene, 
L.~Kofman, A.~D.~Linde and I.~Tkachev,
\emph{``Dynamics of symmetry breaking and tachyonic preheating,''}
Phys.\ Rev.\ Lett.\  {\bf 87} (2001) 011601
[arXiv:hep-ph/0012142].







\bibitem{Berkooz:2002je}
M.~Berkooz, B.~Craps, D.~Kutasov and G.~Rajesh,
\emph{``Comments on cosmological singularities in string theory,''}
JHEP {\bf 0303} (2003) 031
[arXiv:hep-th/0212215].

\bibitem{Felder:2002sv}
G.~N.~Felder, L.~Kofman and A.~Starobinsky,
\emph{``Caustics in tachyon matter and other Born-Infeld scalars,''}
JHEP {\bf 0209} (2002) 026
[arXiv:hep-th/0208019].

\bibitem{Frolov:2002rr}
A.~V.~Frolov, L.~Kofman and A.~A.~Starobinsky,
\emph{``Prospects and problems of tachyon matter cosmology,''}
Phys.\ Lett.\ B {\bf 545} (2002) 8
[arXiv:hep-th/0204187].

\bibitem{Kluson:2003rd}
J.~Kluson,
\emph{``Particle production on half S-brane,''}
arXiv:hep-th/0306002.


\bibitem{Gubser:2003vk}
S.~S.~Gubser,
\emph{``String production at the level of effective field theory,''}
arXiv:hep-th/0305099.


\bibitem{Buchel:2002tj}
A.~Buchel, P.~Langfelder and J.~Walcher,
\emph{``Does the tachyon matter?,''}
Annals Phys.\  {\bf 302} (2002) 78
[arXiv:hep-th/0207235].




\bibitem{Birrell:ix}
N.~D.~Birrell and P.~C.~Davies,
\emph{``Quantum Fields In Curved Space,''}
\href{http://www.slac.stanford.edu/spires/find/hep/www?irn=998621}
{SPIRES entry}

\bibitem{Fulling:nb}
S.~A.~Fulling,
\emph{``Aspects Of Quantum Field Theory In Curved Space-Time,''}
\href{http://www.slac.stanford.edu/spires/find/hep/www?irn=2122588}
{SPIRES entry}

\bibitem{Wald:yp}
R.~M.~Wald,
\emph{``Quantum Field Theory In Curved Space-Time 
And Black Hole Thermodynamics,''}
\href{http://www.slac.stanford.edu/spires/find/
hep/www?irn=3231020}{SPIRES entry}












\bibitem{Kluson:2003sh}
J.~Kluson,
\emph{``The Schroedinger wave 
functional and S-branes,''}
Class.\ Quant.\ Grav.\  {\bf 20} (2003) 4285
[arXiv:hep-th/0307079].

\bibitem{Sen:2003zf}
A.~Sen,
\emph{``Moduli Space of Unstable 
D-branes on a Circle of Critical Radius,''}
arXiv:hep-th/0312003.

\bibitem{Long:1996wf}
D.~V.~Long and G.~M.~Shore,
\emph{``The Schrodinger Wave Functional 
and Vacuum States in Curved Spacetime,''}
Nucl.\ Phys.\ B {\bf 530} (1998) 247
[arXiv:hep-th/9605004].

\bibitem{Guven:1987bx}
J.~Guven, B.~Lieberman and C.~T.~Hill,
\emph{``Schrodinger Picture Field 
Theory In Robertson-Walker Flat Space-Times,''}
Phys.\ Rev.\ D {\bf 39} (1989) 438.


\end{thebibliography}
\end{document}